\documentclass[prb,twocolumn]{revtex4}
\usepackage[dvips]{graphicx}

\begin{document}

\bibliographystyle{apsrev}

\title{Temperature dependence of ambipolar diffusion in silicon--on--insulator}
\author{Hui~Zhao}

\affiliation{Department of Physics and Astronomy, The University of Kansas, Lawrence, Kansas 66045}

\begin{abstract}
Spatiotemporal dynamics of electron--hole pairs locally excited in a silicon--on--insulator structure by indirect interband absorption are studied by measuring differential transmission caused by free--carrier absorption of a probe pulse tuned below the bandgap, with 200--fs temporal and 3--$\mu$m spatial resolution. From sample temperatures of 250 K to 400 K, the ambipolar diffusivity decreases, and is similar to reported values of bulk silicon. Cooling the sample from 250 K to 90 K, a decrease of ambipolar diffusivity is observed, indicating important influences of defects and residual stress on carrier diffusion. No detectable density dependence of ambipolar diffusivity is observed.
\end{abstract}

\maketitle

In a silicon--on--insulator (SOI) structure, a layer of crystalline silicon with a thickness on the order of 1~$\mu$m is grown on an insulating substrate like silicon dioxide or sapphire. Besides being widely used in electronics,\cite{jap934955} it is also an attractive structure for photonics. The large index contrasts at the silicon/insulator and silicon/air interfaces cause strong confinement of light, forming a high--quality waveguide. It also has the potential for integration with SOI--based electronics to achieve integrated optoelectronic devices. In recent years, significant breakthroughs have been made in SOI--based photonics, including demonstrations of Raman lasing\cite{nature433292,nature433725}, optical parametric gain\cite{nature441960}, electro--optic modulator\cite{nature435327,nature441199} and slow light\cite{nature43865,naturephotonics165}.

Ambipolar diffusion is a fundamental process in photonic devices, where carriers are optically excited as pairs of electrons and holes that are bound together by Coulomb attraction, and therefore move in the device as a pair. Such an ambipolar diffusion is therefore the main mechanism to transfer optical excitations in the devices. So far, ambipolar diffusion in SOI has not been studied experimentally. Unlike unipolar transport of electrons or holes, ambipolar transport does not involve any charge movement. Therefore, electric detection techniques that are generally used in charge transport studies are inadequate. Several optical techniques including transient grating,\cite{b307346} pump--probe based on interband absorption,\cite{b385788} and photoluminescence\cite{apl531937,l94137402} have been used to study ambipolar diffusion in direct--bandgap semiconductors and bulk silicon. However, it is difficult to use these techniques to study carrier dynamics in SOI, due to small thickness and indirect bandstructure of the silicon layer.

In this Letter, I report experimental studies of ambipolar diffusion in SOI by a high resolution optical pump--probe technique based on free--carrier absorption (FCA). The technique allows us to directly monitor, in real space and real time, spatiotemporal dynamics of electron--hole pairs with 200--fs temporal and 3--$\mu$m spatial resolution. Ambipolar diffusivities of a sample containing a 750--nm silicon layer grown on a sapphire substrate are measured by monitoring spatial expansion of the carrier density profile after local excitation. From sample temperatures of 250 K to 400 K, the ambipolar diffusivities are similar to reported values of bulk silicon. Cooling the sample from 250 K to 90 K, a decrease of ambipolar diffusivity is observed, indicating important influences of defects and residual stress on carrier diffusion. Furthermore, no detectable density dependence of ambipolar diffusivity is observed in the density range from 0.5$\times 10^{17}$ to 3$\times 10^{17}/\mathrm{cm^3}$.

Figure 1 shows the experimental setup and geometry of the FCA--based pump--probe technique. A 200--fs pump pulse with a central wavelength of 737~nm is obtained by frequency doubling the signal output of an optical parametric oscillator pumped at 80~MHz by a Ti:Sapphire laser. The pump pulse is focused on the sample through a microscope objective to a spot size of 2~$\mu$m full width at half maxima (FWHM). The sample is composed of a 750--nm silicon layer grown on a 350--$\mu$m sapphire substrate along the (100) direction. The silicon layer is p--doped (Boron) with a doping density of $\sim 10^{15}/\mathrm{cm}^3$. With a photon energy (1.682~eV) lower than the direct bandgap (3.40~eV) but higher than the indirect bandgap (1.12~eV) of silicon at room temperature, the pump pulse excites carriers by phonon--assisted indirect absorption (Fig.~1, right panel). After excitation, the carriers diffuse from high density to low density regions, causing expansion of the carrier density profile within the silicon layer.

\begin{figure}
 \includegraphics[width=7.5cm]{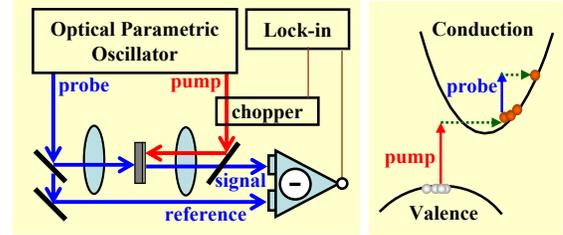}
 \caption{The optical pump--probe system (left panel) and the excitation/probe scheme (right panel).}

\end{figure}

A 200--fs probe pulse with a central wavelength of 1760 nm is obtained from the idler output of the optical parametric oscillator. It is focused to a spot of 3.6~$\mu$m (FWHM) on the sample from the back side. Since the probe photon energy (0.70~eV) is lower than the indirect bandgap of silicon, interband absorption is forbidden. The total absorption coefficient of the probe can be written as $\alpha=\alpha_{0}+\sigma N$, where $\alpha_0$ describes absorptions unrelated to carriers. The $\sigma$ is the cross section of FCA containing contributions of both electrons and holes, with identical density profiles $N$ since they move together in the ambipolar process. The density profile is measured by detecting differential transmission $\Delta T/T_0 \equiv [T(N)-T_0]/T_0$, that is, the normalized difference between the transmissions with [$T(N)$] and without ($T_0$) carriers. It is straightforward to show that, under the condition of $\alpha L<<1$, where $L$ is the sample thickness, $\Delta T/T_0=-\sigma L N$.

Therefore, the spatiotemporal dynamics of carriers can be monitored by measuring $\Delta T/T_0$ as functions of time and space. In order to achieve sufficient signal--to--noise ratio in measuring $\Delta T/T_0$, balanced detection and lock--in detection are combined.\cite{l96246601,b75075305,b72201302} As shown in Fig.~1, a portion of the probe pulse is taken before the sample as a reference pulse, and is measured by one photodiode of a balanced detector. The other portion transmitting through the sample carries signal and is measured by the other photodiode. The balanced detector outputs a voltage proportional to the difference of optical powers on the two photodiodes. The power of the reference pulse is carefully adjusted to match the signal pulse with the pump pulse blocked. The intensity noise of the probe pulse, which is the dominant noise source, is evenly distributed to the two photodiodes, therefore canceled at the output of the balanced detector. With carriers injected by the pump pulse, the power of the signal pulse on the photodiode decreases due to FCA, resulting in a nonzero output voltage of the balanced detector. This voltage signal is proportional to $\Delta T/T_0$ and is measured by a lock--in amplifier referenced to a chopper in the pump arm.

Figure~2 summarizes the spatiotemporal dynamics of carriers measured at a sample temperature of 295~K. The $\Delta T/T_0$ is measured as functions of the probe delay ($\tau$) and the distance between the probe and pump spots ($x$), as shown in Fig.~2a. The peak carrier density is estimated as $2.3 \times 10^{17}/\mathrm{cm}^3$ by using a peak pump fluence of 45~$\mu \mathrm{J/cm^2}$ and a reported absorption coefficient ($2.2 \times 10^3 / \mathrm{cm}$ at 737~nm).\cite{sse2077} By using a reported FCA cross section of electron--hole pairs ($2 \times 10^{-17} / \mathrm{cm^2}$ at 1760~nm)\cite{b273611} and the sample thickness of 750~nm, one expects a peak $\Delta T/T_0=-\sigma L N = 3.45 \times 10^{-4}$, in agreement with the measured value ($3.6 \times 10^{-4}$). As a further verification of the validity of the technique, the peak $\Delta T/T_0$  is measured as a function of the peak pump pulse fluence, which is proportional to the peak carrier density. The observed linear dependence (Fig.~2b) ensures that the $\Delta T/T_0$ truly measures the carrier density without distortion.

\begin{figure}
 \includegraphics[width=7.5cm]{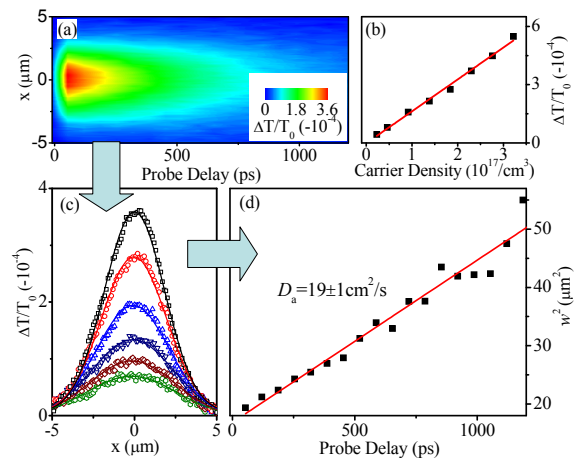}
 \caption{Spatiotemporal dynamics of carriers at 295~K with a peak carrier density is $2.3 \times 10^{17}/\mathrm{cm}^3$.. (a) Differential transmission as functions of probe delay and probe spot position. $x=0$ is defined as where the pump and probe spots overlap.  (b) Peak differential transmission as a function of peak carrier density. (c) Spatial profiles of differential transmission at selected probe delays of (from top to bottom) 50, 190, 390, 590, 790, and 990~ps. (d) Square of profile width $w$ (FWHM) deduced by Gaussian fits as a function of probe delay.}
\end{figure}

To quantitatively study the transport, spatial profiles measured at different probe delays are fit by Gaussian function to deduce the width $w$ (FWHM). A few examples of the profiles with the fits are shown in Fig.~2c. Figure~2d shows the square of $w$ as a function of probe delay. For a diffusion process with initial Gaussian distribution, it is well known that the profile remains Gaussian with the width increasing as\cite{b385788}
\begin{equation}
w^2(\tau)=w^2(\tau_0)+16ln(2)D_a(\tau-\tau_0),
\end{equation}
where $D_a$ is the ambipolar diffusivity. From a linear fit to the data a $D_a=19 \pm 1~\mathrm{cm^2/s}$ is obtained. It is worth mentioning that, due to the finite probe spot size, the profiles measured are actually convolutions of the probe spot and the actual carrier density profiles. However, since both the probe spot and the carrier density profiles are Gaussian, $w^2=w^2_p+w^2_N$, where $w_p$ and $w_N$ are the widths of the probe spot and the carrier density profile, respectively. Since $w^2_p$ exists in both sides of Eq.~1, the convolution doesn't influence the measurement of $D_a$. Furthermore, this procedure of deducing $D_a$ is independent upon the carrier lifetime since the electron--hole recombination only influences the height, not the width, of the profiles.\cite{b67035306}

\begin{figure}
 \includegraphics[width=7.5cm]{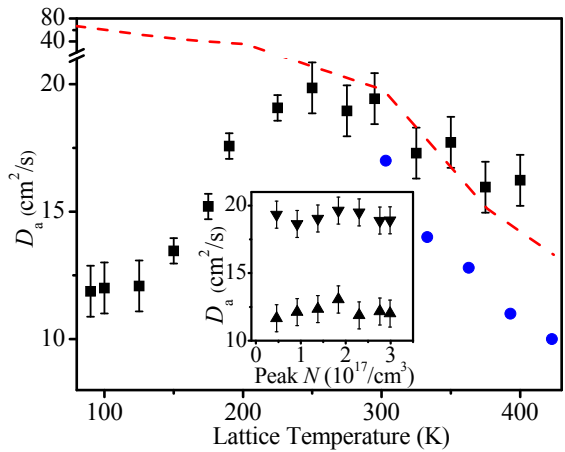}
 \caption{Temperature dependence of ambipolar diffusivity of SOI (squares). The circles are previously reported data on bulk silicon\cite{jap762855}. The dashed line is calculated from known electron and hole mobilities on bulk silicon\cite{sse2077}. Inset: Ambipolar diffusivity as a function of carrier density measured at 90~K (up--triangles) and 295~K (down--triangles).}
\end{figure}

The procedure summarized in Fig.~2 is used to study the ambipolar diffusivity as functions of lattice temperature and carrier density. The results are shown in Fig.~3. The inset shows the dependence of $D_a$ on carrier density measured at 90 and 295~K. No detectable density dependence is observed. This is consistent with the fact that, since carrier--carrier scattering only exchanges momentum among the carriers but conserves the total momentum, it does not directly influence the diffusion. The squares in the main panel of Fig.~3 show the measured ambipolar diffusivity as a function of temperature with a fixed peak carrier density of $2.3 \times 10^{17} / \mathrm{cm^3}$. I find that $D_a$ first increases with temperature from 90 to about 250~K, then decreases with  increased temperature.

Ambipolar diffusivity is determined by the thermal energy of carriers and their interactions with phonons and lattice imperfections. The phonon scattering rate increases with the temperature, but the scattering rate with ionized impurities decreases with temperatures.\cite{Neamenbook} Therefore, complicated temperature dependence of ambipolar diffusion can be expected, and has indeed been generally observed in, e.g. GaAs.\cite{apl531937}

Finally, let us compare the ambipolar diffusivities in SOI and bulk silicon with low doping concentrations. For this purpose, previously reported ambipolar diffusivities in bulk silicon from 300 to 400~K deduced from charge transport measurements are shown in Fig.~3 (circles).\cite{jap762855} I am not aware of any measurement of ambipolar diffusivity in bulk silicon below room temperature. Alternatively, the dashed line in Fig.~3 shows data calculated from generally accepted values of electron and hole mobilities\cite{sse2077} by using Einstein's relation and the well-known relation $D_a=2D_eD_h/(D_e +D_h)$, where $D_e$ and $D_h$ are electron and hole diffusivities.\cite{Neamenbook} Although the phonon scattering rates are similar, the crystal quality of silicon layer grown on sapphire is known to be low, with higher defect density due to lattice mismatch\cite{jap934955} and residual stress caused by different thermal expansion coefficients\cite{apl40895}. In the high temperature regime where phonon scattering dominates, ambipolar diffusivity in SOI is reasonably consistent with bulk silicon deduced from electron and hole mobilities. The previously reported values\cite{jap762855} are generally smaller but show similar temperature dependence. However, a significant difference is observed when temperature is lowered from 250 to 90~K. The decrease of ambipolar diffusivity in SOI with temperature suggests strong influences of defects and stress on carrier diffusion in this regime.

In summary, an optical pump--probe technique based on free--carrier absorption with high spatial and temporal resolution is developed and applied to study spatiotemporal dynamics of carriers in silicon--on--insulator directly, in real space and real time. Ambipolar diffusivity is measured in a temperature range from 90 to 400~K and in the carrier density range from 0.5$\times 10^{17}$ to 3$\times 10^{17}/\mathrm{cm^3}$. The study provides fundamental parameters for developing SOI--based photonics, and demonstrates a technique that can be generally applied to study carrier dynamics in SOI.

I acknowledge helpful discussions with Carsten Timm and Karl Higley, and financial support from GRF of KUCR.

\end{document}